# ENTROPY AND ESCHATOLOGY: A COMMENT ON KUTROVÁTZ'S PAPER *HEAT DEATH IN ANCIENT AND MODERN THERMODYNAMICS*

Milan M. Ćirković

*Astronomical Observatory Belgrade*
*Volgina 7, 11160 Belgrade, SERBIA, Yugoslavia*

*e-mail:* `arioch@eunet.yu`

# ENTROPY AND ESCHATOLOGY: A COMMENT ON KUTROVÁTZ'S PAPER *HEAT DEATH IN ANCIENT AND MODERN THERMODYNAMICS*

**Abstract.** Recent intriguing discussion of heat death by Kutrovátz is critically examined. It is shown that there exists another way of answering the heat death puzzle, already present in the ancient philosophical tradition. This alternative route relies not only on the final duration of time (which has been re-discovered in modern times), but also on the notion of observational self-selection, which has received wide publicity in the last several decades under the title of the anthropic principle(s). We comment here on some further deficiencies of the account of Kutrovátz. Although the questions Kutrovátz raises are important and welcome, there are several errors in his treatment of cosmology which marr his account of the entire topic. In addition, the nascent discipline of physical eschatology holds promise of answering the basic explanatory task concerning the future evolution of the universe without appealing to metaphysics. This is a completely novel feature in the history of science, in contradistinction to the historical examples discussed by Kutrovátz.

## 1. Introduction

The problem of heat death—and the general issue of the thermodynamical temporal asymmetry—is one of the long-standing and ever-inspiring topics in the history of human philosophical and scientific thought, comparable to Zeno's paradoxes on space and motion or the Hicetas-Berkeley-Mach puzzle of the origin of inertia of matter. Recent discussion of the topic by Kutrovátz (2001) in this Journal is indicative of the great surge of interest in physical, philosophical as well as the historical aspects of this and related themes (for some other recent discourses, see Price 1996, 2002; Lieb and Yngvason 1999; Albert 2000; Uffink 2001; Uffink and Brown 2001). It brings (i) a very welcome appraisal of the state of the problem in by far the most influential ancient physical theory—Aristotelian physics—and (ii) unearths some of the valuable comparison between that state of affairs and the situation in modern cosmological thought. However, its treatment is incomplete in several important points, and contains—especially it the latter part—some (widely spread) misconceptions and mispresentations.

Our goal in this note is, therefore, to correct some omissions in Kutrovátz's manuscript, as well as to suggest generalizations of its conclusion in the light of the nascent astrophysical discipline of physical eschatology. This seems an excellent case study for emphasizing the degree our cosmological knowledge is determined by the underlying information flow. In addition, we point out another, surprisingly modern, answer to the heat death puzzle which existed in the antiquity, in a tradition different from the Aristotelian one. We begin with the latter topic.

**2. Alternative answer to the heat death problem**

In contradistinction to the Aristotelian "prime movent" which is eternally co-present with the world, the modern evolutionary idea that the universe is of finite age and is currently unwinding as a consequence of finite amount of primordial motion was also present in antiquity, even prominent in some circles. Unfortunately, those were the circles of philosophers of nature who were either agnostics or open atheists, and therefore were carefully supressed and neglected in subsequent centuries. The elements of such a worldview were present in cosmologies of Anaximandros, Empedocles, Anaxagoras and Democritus, as well as in some later epigones. We mention here briefly some of these ideas.

In the very first chapter of the immortal history of Thucydides, there is a famous statement that before his time—i.e. about 450 BC—nothing of importance (συ μεγαλα γευεσθαι) had happened in history. This startling statement has been correctly called "outrageous" by Oswald Spengler, and used to demonstrate the essentially mythological character of the ancient Greek historiography (Spengler 1918; see also Cornford 1965). It may indeed be outrageous from the modern perspective, but it does motivate a set of deeper questions, ultimately dealing with cosmology. The fact that Thucydides did not know (or did not care to know) previous historical events does not change the essential perception of **finiteness** of human history inseparable from the Greek thought. This property starkly conflicts with the notion of an eternal continuously existent world, as it was presented in both modern and ancient cultures. Obviously, it is irrelevant which exact starting point we choose for unfolding historical events. In any case, the number of these events is finite, and the time span considered small even compared to the specific astronomical timescales (some of which, like the precession period of equinoxes, were known in the classical antiquity, as is clear from the discussion in *Timaeus*), not to mention anything about a past temporal infinity. Although there was no scientific archaeology in the ancient world, it was as natural then as it is now to expect hypothetical previous civilizations inhabiting Oikumene to leave some traces—in fact, an infinite number of traces for an eternally existent Oikumene! There are indications that pre-Socratic thinkers have been aware of the incompatibility of this "Thucydidean" finiteness of historical past with the eternal nature of the world. We have already mentioned the solution (periodic singular states) proposed by Empedocles himself. Even earlier, in the fragmentary accounts of the cosmology of Anaximandros, one may note that he proposed an evolutionary origin of humankind in some finite moment in the past, parallel with his basic postulate of separation of different worlds from *apeiron* and their subsequent returning to it.[1] In Anaxagoras' worldview, there is a famous tension between the eternity of the world's constituents and the finite duration of **movement** (and, therefore, relational time) in the world. In the same time, it seems certain that Anaxagoras, together with Anaximandros and Empedocles, was an early proponent of the evolutionary view, at least regarding the origin of humankind (Guthrie 1969).

Finally, an almost modern formulation of the anthropic argument for the finite past has been made in Roman times by Lucretius, who in Book V of his famous poem *De Rerum Natura* wrote the following intriguing verses:

> Besides all this,
> If there had been no origin-in-birth
> Of lands and sky, and they had ever been
> The everlasting, why, ere Theban war

---

[1] This is clear, for instance, from the fragment A 10 in Diels (1983), preserved by Plutarch, in which it is explicitly asserted that formation and destruction of many worlds occurs within the global temporal infinity. In the continuation of the very same excerpt from *Stromateis*, an evolutionary doctrine is attributed to Anaximandros: "...Farther he says that at the beginning man was generated from all sorts of animals, since all the rest can quickly get food for themselves, but man alone requires careful feeding for a long time; such a being at the beginning could not have preserved his existence." (Fairbanks 1898) Hyppolites quotes Anaximandros as emphasizing the nature of *apeiron* as eternal (B 2), obviously in opposition to mankind, which has a fixed beginning in time. Even more intriguing is the doctrine ascribed to Anaximandros by Cicero: "It was the opinion of Anaximandros that gods have a beginning, at long intervals rising and setting, and that they are the innumerable worlds. But who of us can think of god except as immortal?" Did he have in mind essentially what we today in SETI-related discussions denote as supercivilizations (e.g. Barrow & Tipler 1986)?

> And obsequies of Troy, have other bards
> Not also chanted other high affairs?
> Whither have sunk so oft so many deeds
> Of heroes? Why do those deeds live no more,
> Ingrafted in eternal monuments
> Of glory? Verily, I guess, because
> The Sun is new, and of a recent date
> The nature of our universe, and had
> Not long ago its own exordium.[2]

For highly scientific-minded Lucretius, the shortness of human history **is** very strange on the face of hypothesis of the eternal existence of the world. Although the references to "eternal monuments" and "other bards" may sound naive, it is clear that he had in mind any form of transmission of information from the past to the present; and an infinite amount of information from an infinite past. His empirical assessment of the surrounding world clearly shows the absence of such information. Therefore, an explanation is needed. The simplest explanation, as Lucretius was highly aware, is to treat the argument as *reductio ad absurdum* of the starting hypothesis (eternal nature of the world) and to assume that the world is of finite—and relatively small—age.

The depth of Lucretius' thought in this passage is almost amazing, especially when the historical blindness of subsequent generations to this same argumentation is taken into account. The Lucretius' argument applies to the classical Newtonian universe of infinite age, as well as to modern stationary alternatives to the evolutionary cosmology (like the classical steady-state theory). It emphasizes the **technological** nature of possible evidence ("ingrafted in… monuments"). This is exactly what modern cosmologists Davies and Tipler have had in mind when constructing the anthropic argument in order to refute the eternal cosmologies of our epoch. Lucretius' monuments play essentially the same role as Tipler's von Neumann probes sent by advanced intelligent communities (Tipler 1982). Their absence testifies on the finite past.

We conclude that a rather modern idea of the (relatively) recent origin of the universe and its "running down" has been present in classical antiquity. Of course, it did never attain the attention and considerations accorded to the Aristotelian picture, but it has been present nevertheless, testifying upon the high degree of scientific sophistication of the ancient world. It has been resurrected in its modern form, for instance, in the famous debate on the thermodynamical recurrence between Boltzmann and Zermelo (cf. Steckline 1983), which is the first instance scientific (in the modern sense) cosmological speculations appeared in a respected peer-reviewed research journal. In the course of XX century, it has been incorporated into the standard cosmological lore, in particular after the victory of evolutionary models over their steady-state rival in the "great battle" of 1950s and 1960s (Kragh 1996). It is exactly this lore we now turn to.

### 3. Modern cosmology and some errors in Kutrovátz's account

In this Section we list and briefly consider several important astrophysical points missing in Kutrovátz's account of the modern answer to the heat death puzzle. These considerations will certainly help in highlighting the entire scope of the problem whose important aspects, we wholeheartedly agree with Kutrovátz, are still open.

### 3.1. Gravitational field as the major source of entropy

The basic missing part in Kutrovátz's toy model of the entropy evolution of the universe is gravitational entropy. It has very slowly dawned of physicists and cosmologists that, apart from the thermodynamical entropy, gravitational field may store a huge quantity of the internal degrees of freedom, thus having potentially huge entropy in itself. This has been realized in the black hole context in early 1970s, with the revolutionary studies of Bekenstein (1973) and Hawking (1974). Bekenstein-Hawking formula gives us a hold on the entropy of gravitational field of a black hole, and there certainly are ways of discerning gravitational entropy in other cases, although they are still highly speculative (being dependent, of course, on the structure of correct quantum theory of gravitation!). Introducing gravitational entropy enables solving several important puzzles in astrophysics and cosmology (e.g. Penrose 1989). The most important of them is the state of thermodynamical equilibrium of very early cosmological epochs, as discerned from the high isotropy of the cosmological black-body radiation. If the classical thermodynamical entropy is all there is, how then could anything occur in the universe after the epoch of recombination at redshift $z \sim 1500$? We would naively expect plasma to simply reach complete equilibrium and remain in such state forever. Instead, gravitational clumping acted to reduce thermodynamical entropy at the expense of gravitational entropy which greatly increased during the process of cosmological structure formation. And cosmological structure formation, in turn, enabled all the wealth of subsequent physical, chemical and biological processes we are now dealing with in sciences; it is, among other things, the cause of our appearance on Earth as intelligent observers.

---

[2] In translation of William E. Leonard, available via WWW Project Gutenberg (Lucretius 1997).

**3.2. Thermodynamical capacities of gravitating systems**

When Kutrovátz writes that "we can see here that the internal energy is not constant because it is tranformed into the energy of the gravitational field: this is a peculiar feature of the universe as a thermodynamic system", he is only partially correct. While this undoubtedly holds true for the universe, it is also true for many other astrophysical systems, notably stars and galaxies. In fact, it is true for any system that is held together by a long-range force like gravity. Suppose that we add energy to a star (say by a giant laser beam), and then wait some time for the relaxation processes to occur. We shall notice that its size has increased and its temperature will actually decrease (familiar example of the negative heat capacity).

This may seem irrelevant for the particular cosmological case. However, the conclusion often drawn from the same property when applied to the Hubble expansion, namely that the Hubble expansion is the cause of the entropy gradient (i.e. that expansion creates new extropy), is wrong. This point has been the focus of a fierce controversy raging in 1960s and 70s, beginning with the classical study of Thomas Gold (1962; see also Davies 1974; Layzer 1976). The idea was to explain the departure from thermodynamical equilibrium assuming that the universal expansion creates new states for new configurations of matter, so that the entropy of matter begins to lag more and more behind the **maximal** entropy possible. Ingenious as it was, this idea has been abandoned since for several reasons, the main being that the "special" low entropy nature of the initial Big Bang singularity is so exceptional (as calculable in principle from Boltzmann formula $S = k \ln W$; see Penrose 1989) that the amount of subsequent expansion produces almost negligible effect. In other words, we still need to explain the very low initial **gravitational** entropy, whose subsequent increase more than offset the apparent (thermodynamical) entropy decrease after the recombination epoch.

**3.3. Open vs. closed universes and cosmological constant**

In several places in Kutrovátz (2001) the confusion of open/closed vs. ever-expanding/recollapsing is perpetuated. This has become very widespread as a consequence of general neglect (in particular in textbooks) of models with cosmological constant (or any other form of "dark energy") in the approximate 1930-1990 period. Today, after the spectacular results of recent cosmological observations of distant supernovae, as well as of anisotropies of the microwave background radiation, we are in much better position. For a nicely written summary of the situation at present see Krauss and Turner (1999).

Let us summarize some cosmological basics. $\Omega$ is the ratio of physical density of all matter fields to the so-called critical density necessary for universe to stop current expansion and recollapse toward Big Crunch. In the misleading textbook discourse which Kutrovátz uncritically accepts, $\Omega \leq 1$ universes will expand forever, while $\Omega > 1$ universes will recollapse. This strictly applies only to the case of matter fields possessing "regular" equation of state; in the presence of vacuum energy, indicated by recent cosmological supernovae experiments, the situation becomes more complicated. There, we can write $\Omega = \Omega_m + \Omega_\Lambda$, $\Omega_\Lambda$ being the contribution of dark energy. Now, even $\Omega > 1$ (= topologically closed) universes may expand forever, under the condition that the sign of dark energy is positive (corresponding to the repulsive effective force). This will occur for any $\Omega_m \leq 1$, and if $\Omega_m > 1$ for (e.g. Carroll, Press and Turner 1992)

$$\Omega_\Lambda \geq 4\Omega_m \left\{ \cos\left[\frac{1}{3}\arccos\left(\frac{1-\Omega_m}{\Omega_m}\right) + \frac{4\pi}{3}\right] \right\}^3 .$$

Thus we have a degeneracy in relationship of topological properties (open vs. closed) and its dynamical future (ever-expanding vs. recollapsing). With positive cosmological constant (which seems realistic; e.g. Perlmutter et al. 1999; Riess et al. 2001; Tucci, Contaldo & Bonometto 2001) topologically closed universe may also expand forever, especially if its contribution is so large as indicated by observations ($\Omega_\Lambda \approx 0.7$). We have no observational way of determining whether the real universe is open or closed, since—and especially if one believes in inflationary models—the total cosmological density is $\Omega = 1 \pm \varepsilon$, $\varepsilon$ being of the order of $10^{-5}$. However, we may be virtually certain that the universe, no matter its topological properties, will expand forever.

**3.4. Extropy in open universes and the definition of the heat death**

If the definition of heat death is—as classically always assumed—the state of **reached** maximal entropy of a closed physical system, than the issue whether heat death will occur in an ever-expanding universe is not trivial or obvious. Although entropy may increase forever, this does not mean it will reach maximal value, if that maximal value can recede indefinitely. In fact, it seems that for the classical Einstein-de Sitter universe (matter-dominated universe with $\Omega = \Omega_m = 1$) exactly that is the case: in one of the first studies of cosmological future (appropriately entitled "Entropy in an Expanding Universe"), Frautschi (1982) wrote:

> It is apparent... that the entropy in a causal region falls steadily further behind max S during most of the cosmic history. $S/S_{max}$ does increase temporarily during the period of stellar and galactic black hole formation. Life as we know it develops during the same period, utilizing the much smaller, but conveniently arranged entropy generation on a planet or planets situated near nucleosynthesizing stars. ...the expanding universe does "die" in the sense that the entropy in a comoving volume asymptotically approaches a constant limit.

This is **not** heat death as classically understood (e.g. Eddington 1931). In the similar vein, Adams and Laughlin (1997) in the comprehensive study of almost all aspects of physical eschatology notice that

> Thus far in this paper, we have shown that entropy can be generated (and hence work can be done) up to cosmological decades η ~ 100. [Cosmological decades are defined as epochs of time t = $10^n$ years] ...The formation of larger and larger black holes, can continue as long as the universe remains spatially flat and the density perturbations that enter the horizon are not overly large. The inflationary universe scenario provides a mechanism to achieve this state of affairs, at least up to some future epoch... Thus the nature of the universe in the far future η » 100 may be determined by the physics of the early universe (in particular, inflation) at the cosmological decade η ~ –45.

But it is not necessary to enter into wealth of technical details of these and related studies here; it is enough to point out that there already exists a sizeable volume of scientific literature on the entropy production in the cosmological future—literature Kutrovátz unfortunately ignores.

On the other hand, in the realistic models with the positive cosmological constant, heat death may be operationally defined as the state of reached maximal entropy **within the event horizon**. In this case the assertion of Kutrovátz that the heat death is inescapable is correct.

## 4. Discussion: physical eschatology and open systems

Physical eschatology is a rather young branch of astrophysics, dealing with the future fate of astrophysical objects, as well as the universe itself. Landmark studies in physical eschatology are those of Rees (1969), Dyson (1979), Tipler (1986) and Adams and Laughlin (1997). Some relevant issues have been discussed in the monograph of Barrow and Tipler (1986), as well as several popular-level books (Islam 1983; Davies 1994; Adams and Laughlin 1999). Since the distinction between knowledge in classical cosmology and physical eschatology depends on the distinction between past and future, several issues in the physics and philosophy of time are relevant to the assessment of eschatological results and *vice versa*.

At a first glance, it seems that—in contradistinction to most other physical disciplines—physical eschatology is at least fortunate enough to deal with the exemplary closed system: the universe itself. But there is one crucial sense in which physical eschatology is—contrary to this impression—a study of open systems. This is nicely manifested in the discussion of heat death of the universe. A necessary ingredient in most serious discussions of physical eschatology is presence of living and intelligent systems in future of the universe (which *ex hypothesi* did not exist in its past). Dyson has been the first to boldly spell it out in 1979:

> It is impossible to calculate in detail the long-range future of the universe without including the effects of life and intelligence. It is impossible to calculate the capabilities of life and intelligence without touching, at least peripherally, philosophical questions. If we are to examine how intelligent life may be able to guide the physical development of the universe for its own purposes, we cannot altogether avoid considering what the values and purposes of intelligent life may be. But as soon as we mention the words value and purpose, we run into one of the most firmly entrenched taboos of twentieth-century science.

Future of universes containing life and intelligence is **essentially** different from the future of universes devoid of such forms of complex organization of matter; as well as different from the past of the same universes in which complexity was lower. In a similar vein, John A. Wheeler wrote in a beautiful paper on the relationship of quantum mechanics and cosmology (Wheeler 1988):

> Minuscule though the part is today that such acts of observer-participancy play in the scheme of things, there are billions of years to come. There are billions upon billions of living places yet to be inhabited. The coming explosion of life opens the door to an all-encompassing role for observer-participancy: to build, in time to come, no minor part of what we call *its* past—*our* past, present and future—but this whole vast world.

Taking into account intentional actions of intelligent beings transforms this field into a study of systems open to the interaction with such advanced noospheres (particularly in light of our virtually complete ignorance of the physics of conscience and intelligence at present). The pioneering study of Dyson quoted above presents an excellent example (although it may be wrong in quantitative details; see Krauss and Starkman 2000) of such a research: in it Dyson envisages a way for advanced intelligent communities to avoid heat death via ingenious methods of energy conservation and information flow optimization. Dyson's work provoked a lot of subsequent research activity. Although we cannot enter into this discussion here, the general moral that future of the universe, including the possible heat death, has recently entered the physical—instead of metaphysical—domain seems inescapable. This stands in a stark contrast to the course of thinking about the cosmological future in previous epochs which Kutrovátz describes, and heralds a completely new era in our understanding of the universe.

**Acknowledgements.** The author wishes to acknowledge Prof. Branislav Nikolić and Vesna Milošević-Zdjelar for their kind help in finding several key references. Special thanks are owed to Nataša Bulut for her invaluable inspiration and wholehearted support.